\centerline{\bf ON M THEORY, QUANTUM PARADOXES AND THE NEW RELATIVITY}
\bigskip
\centerline{Carlos Castro}
\centerline{Center for Theoretical Studies of Physical Systems}
\centerline{Clark Atlanta University}
\centerline{Atlanta, GA. 30314}
\smallskip

\centerline{Alex Granik}
\centerline{Department of Physics }
\centerline{University of the Pacific}
\centerline{Stockton, California 95211}
\smallskip

\centerline{January 2000}
\smallskip

\centerline {(Dedicated to the memory of Leonard Ainsworth, a true pioneer and a scientist )} 

\bigskip

\centerline{\bf Abstract}
\bigskip

Recently a New Relativity Principle has been proposed by one of the authors as the underlying physical and geometrical foundations of String and {\bf M} Theory. 
It is explicitly shown that within the framework of the New Relativity Theory, some Quantum Mechanical Paradoxes like the Einstein-Rosen Podolsky and 
the Black Hole Information Loss, are easily resolved. Such New Relativity Theory requires the introduction of an    
Infinite Dimensional Quantum Spacetime as has been shown recently by one of us. This can be viewed as just another way of looking at  
Feynman's path integral formulation of Quantum Mechanics. Instead of having an infinite dimensional funcional integral over $all$  
paths, smooth, forwards and backwards in time, random and fractal, 
in a finite-dimensional spacetime, one has  a finite number of paths in an Infinite Dimensional Quantum Spacetime.   
We present a few-lines proof why there is no such a thing as an {\bf EPR}  Paradox in 
this New Relativity theory. The reason is {\bf not}  due to a superluminal information speed but to a {\bf divergent}  information charge density. 
In the infinite dimensional limit, due to the properties of gamma functions, the hypervolume enclosed by a $D$-dim hypersphere, of finite nonzero radius,  shrinks to 
zero : to a {\bf hyperpoint} , the infinite-dimensional analog af a point.     
For this reason, Information flows through the infinite-dimensional hypersurface of  nonzero radius, but zero size, the hyperpoint, in an instant.   
In this fashion we imbue an abstract mathematical "point"  with a true physical meaning : 
it is an entity in infinite dimensions that has zero hypervolume at nonzero radius . 
A plausible resolution of the Information Loss Paradox in Black Holes is proposed.

\bigskip

\centerline{\bf 1. Introduction : Historical background }
\bigskip
From the very beginning the relativity principle has  been one of the cornerstones of mechanics. In {\bf 1686} , in the 
opening pages of $Principia$, Newton wrote : " I do not define time, space, place, and motion, as being well known to all. Only 
I must observe, that the common people conceive those quantities under no other notions but from the relations they bear to sensible objects ". 
Based on this preamble, Newton introduced the absolute space, time, and motion which in his own words do not bear " relation to anything external " .
It is clear that such absolute space, time, motion, are purely metaphysical notions, stand outside the realm of physics, and serve the purpose of 
" geometrization " of physical phenomena. Thus at the foundations of Newtonian mechanics lie the above metaphyisical concepts. 
The very same Metaphysical concepts that so many members of the scientific community, by their own admission, are unwilling to accept.    

Much later, Mach rather broadly , and then Einstein definitely,  set Physics on physical ground by defining measurements 
relative only to the physical 
phenomena, and not to the metaphysical entities.  
Instead of vaguely-defined ( if at all) metaphyiscal concepts of absolute space, time, and motion, new 
( rather narrow) but well defined 
physical concepts of physical  measurements  had been introduced. The old "absolutes " had been dismissed by Einstein's Theory of Relativity.     
Measurements required a universal standard which had been furnished by a physical quantity : the speed of light. 

Based on this historical perspective we introduce the New Relativity Principle [1] that rest on the four postulates presented in the next section.  
In the last sections an amazingly simple proof is presented why there is no {\bf EPR} Paradox in such New Relativity Theory. In addition, a plausible 
resolution of the Information Loss in Black Holes is proposed. 
\bigskip
\centerline{\bf 2. The Program of the New Relativity Principle : The Demolition of Today's Absolutes } 
\bigskip

Recently one of the authors has proposed that a New Relativity principle may be 
operating in Nature
which could reveal important clues to find the origins of $M$ theory  
[1]. 
We were forced to introduce this new Relativity principle, where all 
dimensions and
signatures of spacetime are on the same footing, to find a fully 
covariant formulation
of the $p$-brane Quantum Mechanical Loop Wave equations. This New 
Relativity Principle,
or the principle of Polydimensional Covariance as has been called by 
Pezzaglia,
has also been crucial in the derivation of Papapetrou's equations of 
motion of a
spinning particle in curved spaces that was a long standing problem
which lasted almost 50 years  [2]. A Clifford calculus was used where 
all the
equations were written in terms of Clifford-valued multivector 
quantities;
i.e one had to abandon the use of vectors and tensors and replace them 
by
Clifford-algebra valued quantities, matrices, for example .

The New Relativity Theory [1] rests on four  postulates :  

{\bf 1}. The old Bootstrap Idea of Chew : Each $p$-brane is made of {\bf all } the others. To view a single $p$-brane as a  fundamental identity is a meaningless concept.
$p$-branes are defined only in {\bf relation} to others. This is Mach's  principle once again. For this reason, one must include all dimensions and signatures 
on the same footing.
Pezzaglia [2] has called this the principle of Polydimensional Covariance or Dimensional Democracy. The New Relativity theory reshuffles, for example, a string history for a 
$5$-brane history; a $9$-brane history for a membrane history; an $11$-brane history for a history encompassing all other $p$-branes...and so forth. 
Point and extended instantons and tensionless $p$-branes are also inlcluded. The tensionless $p$-brane history excitations of the 
infinite dimensional Quantum Spacetime are the " photons " in this New Relativity Theory. 

We honestly believe that $M$ Theory does {\bf not } stand for mystery, membrane, matrix, master, mother, murky, Moyal....it stands for Mach. 
The New Relativity theory is based on the ultimate 
Machian view of the Quantum Universe ( the Ultimate Machian " Quantum Computer "  ) : 
Relationships among entities are the only meaningful statements one can make . A perfect example of this are : 
spin-networks, quantum-networks, quantum sets, cellular networks, $p$-Adic Physics,...etc.  
Since it is undesirable to run off the letters of the alphabet, by keep adding letters like 
$M, F, S...$ Theory, we gather courage to say that by abandoning the Egocentric 
Anthropomorphic  view of the Universe and, 
instead, 
embracing Mach's view that everything in the Quantum Universe is interconnected, one reaches the end of the alphabet at $Z$ theory : 
$Z$ stands for the ultimate Machian view of the Quantum Universe (
that dismisses the egocentric view of the Universe )
for a $Zenthropic$ Quantum Universe.  
Who are {\bf we}  to say  that we know everything that an electron , a
 photon, a quark  really " sees " ?  
Do  electrons, quarks....perform Feynman
diagrams ? Has anyone seen a point ?  
 
{\bf 2}. Laurent Nottale's Scale-Relativity theory [3]. 

In the final analysis, Physics involves, and is about, 
measurements. Physics {\bf is} an experimental science. Physics deals with experiences. 
On the other hand to  measure something one needs a standard of measurement to compare
measurements with. It is essential , it is of prime importance, to introduce resolutions in Physics. It is meaningless to say that the one has a field at 
{\bf x}. 
By {\bf x} meaning : 
specifying the value of the real number $x$ to an infinite number of non-periodic decimal places. 
In mathematics we can infinitely 
increase the accuracy ( or degree of resolution) at will of any real number by adding digits. However, in practice we cannot 
have such arbitrary
accuracy provided by mathematical constructions. Even writing an infinite aperiodic decimal fraction would require an infinite amount of memory.
Therefore, in Physics, it is necessary to have a finite
universal physical "yardstick" which would define the ultimate Physical Resolution. Nottale's Scale Realtivity
takes the Planck scale as such Universal physical standard of measurement that is invariant, by definition,  under Scale Relativistic transformations of resolutions, 
like the speed of light was in Einstein's Relativity.     

$p$-adic numbers and $p$-adic Physics is a nice attempt to
eliminate the problem of having to specify a real number up to infinite digits [11]. 
The Planck scale is therefore taken as that universal standard of measure invariant under Scale Relativistic transformations of {\bf resolutions} .  
In the same vein that the speed of light was taken by Einstein as the 
maximum speed in Nature, the Planck scale is taken to be the minimum length. 
The speed of light allowed Einstein to embrace space with time, since space and time have different units. 
By the same token, to embrace all dimensions one needs a Universal length scale in {\bf all} dimensions : the Planck Scale. 

As the years pass by, more and more planets have been found confirming Nottale's predictions within his framework of Scale-Relativity. Instead of 
being properly rewarded with increased curiosity and interest in his remarkable theory, he has been increasingly rewarded with insults and a suffocating
censorship [12]. 
As the number of his planet confirmations increases, so does the number of insults increases and the censorship of his work is tightened further. 
Unfortunately, the New Relativity Theory will never be able to explain such odd phenomena.

The Universal scale, in units of $\hbar =c=1$, in {\bf any} dimensions, $D>2$ ( in two dimensions the Einstein-Hilbert action is a topological invariant)  is :

$$\Lambda = G_D ^{ {1\over D-2}}= G_{D-1} ^{{1\over D-3}}  =...........=10^{-33} cms. \eqno (1a)$$
where $G_D, G_{D-1}.....$ are the Newton gravitational coupling constants in different dimensions. 
In the same fashion that in Newtonian Physics one only can assign a  definite
meaning to the {\bf ratio} of masses ( it is meaningless to say that one has a value of $m$ without a comparison to another mass ), in the 
New Relativity theory it only makes sense to write :

$${D-3 \over D-2 } ={ ln G_{D-1} \over ln G_D}. \eqno (1b)$$

In the New Relativity Theory it is meaningless to talk about such things as " compactification " , " decompactification " used 
in the literature that relates the Newton constants in different
dimensions through the small radius of a compactified unseen dimension at low energies.  
It as meaningless as saying that the velocities of the gas molecules in a room experience a dynamical or spontaneous " compactification " to a fixed average value. 

Problems with the compactification picture of Superstring theory from  {\bf 10}  to {\bf 4}  were already alarming signals that something could be wrong. 
Billions and billions of possible
four-dimensional phenomenological theories of the world were obtained : the so-called {\bf uniqueness}  of string theory went out the window when this was found. 
String theory wasn't the problem, assuming a fixed dimensions was ! Witten already proved long ago that something might be inherently wrong with the compactification 
schemes, when he 
showed using Index Theory arguments, that the standard {\bf 11}-dim Supegravity 
Kaluza-Klein compactifications of ordinary manifolds
did not yield chiral fermions in {\bf 4} dimensions. This problem was bypassed in the second string revolution by saying that orbifold compactification were fine because 
orbifolds are not really ordinary manifolds, so things were satisfactory after all. 
The New Relativity Theory does not have to face these challenges. One has a truly infinite-dimensional Quantum Universe which suggests that Topological Field Theories 
could be the most natural candidates for a theory of the world. Since below the Planck scale there is no such thing as a distance; 
it is very likely that Topology should play a more important role.

Conformal Field Theories and their Higher Conformal spin extensions are the ones to use in $D=2$. In $D=2$, one has induced gravity : $W_2, W_3...W_\infty$ gravity as a 
result of integrating out the conformal matter field fluctuations. This replaces the topological invariant Einstein-Hilbert action. 
In $D=1$ dimension there is only extrinsic curvature. One can view a one-dim loop as the boundary of a two-dimensional surface. This allowed [7] to write 
down a String Representation of Quantum Loops from a covariantized phase space Schild action path integral. The effective action for the boundary, 
with induced extrinsic curvature terms was obtained,  in addition to the Polyakov Bulk partition function and the holographic boundary Eguchi wave functional  as well.

{\bf 3}. Noncommutative {\bf C}-spaces. One of the authors  was forced to enlarge the naive notion of commuting spacetime coordinates 
to fully covariantize the Quantum Mechanical Loop
Equations for $p$-branes. One achieved that goal if one extended the notion of ordinary spacetime vectors and tensors, to a Noncommutative {\bf C}-space, 
or Clifford manifold,  where all $p$-branes were unified in one single footing by using Clifford-algebra valued {\bf multivectors}  quantities ( matrices) 
instead of ordinary vectors and tensors. In order to combine objects of 
different dimensionality one needs a length scale : the Planck scale.  

There was a one-to-one correspondence between a {\bf nested} hierarchy of point, loop, {\bf 2}-loop, {\bf 3}-loop .......{\bf p}-loop histories in $D$ dimensions 
encoded in terms of hyper-matrices and single lines in Clifford Manifolds. This is roughly similiar to the aim of Penrose's twistor progam. 
By using Clifford-algebra valued {\bf multivectors}, one could argue why it may be meaningless to say that the cosmological constant is a constant in its 
definition ! The so-called cosmological constant is observer-dependent in this New Relativity Theory : it is just one of the many components of the Clifford multivectors.
Due to Polydimensional Covariance, only the {\bf norm} of such multivector is truly an invariant. So using this simple argument one of us was able to argue why 
it is meaningless to try to measure such constant, unless one is specifying what is the frame of reference one lives in !    

The reader may say that the value of $p=-1$ was not included here. 
Point and Extended Instantons can also be treated very naturally in this framework [1]. The New Relativity Theory reshuffles, for example, a loop-history 
represented by the coordinates : $x^\mu ,\sigma_{\mu\nu}, A $ in one frame of reference, to another history, in another frame of reference, 
represented by the loop-instanton 
$x'^\mu , (\sigma_{\mu\nu} )' , A'=0$. The $x_{CM}$ are the center of mass coordinates of the loop. $A$ is the areal-time spanned by the motion of the loop through spacetime.  
$\sigma_{\mu\nu} $ are the holographic coordinates of the loop. It can reshuffle a massive point history ( a line ) : $x^\mu, \tau \not=0 $ to a 
massive point-instanton  : $x'^\mu, \tau '=0$ in another frame of reference. An so forth.  

{\bf 4}. Quantum Spacetime must be treated from a Multivector-Multiscale point of view. The use of Clifford-valued multivectors was explained above. 
The multi-scale or resolution aspects are  based on Nottale's fractals and El Naschie' s Cantorian-Fractal Quantum Spacetime views  that 
dimensions are {\bf resolution} dependent concepts and not {\bf fixed} notions [3,4]  .  

Nottale, by abandoning the hypothesis of the differentiablity of spacetime , was led to three effects ( at least ) :
{\bf (i) }. The number of geodesics becomes infinite. This forces upon us to jump to a statistical fluid-like description. 
{\bf (ii)} Each geodesic becomes a fractal curve of  higher and higher fractal dimensionality as the resolution of the  " physical 
apparatus " becomes finer  and finer, asympotically approaching the minimum Planck scale resolution where the fractal dimensionality
becomes infinite. This forces us to embed the fractal geodesics in an spacetime of infinite-Hausdorff dimensions.
{\bf (iii)}. The symmetry $dt \rightarrow -dt$ is broken by the non-differentiablity which leads to a {\bf two}-valuedness character of the average
velocity vector and  which is, in Nottale's view,  
the underlying {\bf reason}  why the wave function in QM is {\bf complex}. 

This is not the ultimate status of things. To be consistent and to move forward along the path charted by Mach and Einstein, 
one cannot, and should not , accept this status quo  as the " end of the road " in Physics. This reminds us of the status of things at the end of 
the {\bf 19} century when " two clouds " were the only obstacles hovering over the horizon that prevented the " end " of Physics. 
In fact, one cannot but to feel compelled to say 
that from the beginning, a 
 truly quantum mechanical description of the world must start by abandoning the $very~ notion~ of~ spacetime~ itself$  and other " idols " from our minds, 
as Finkesltein has pointed out [8]. This is precisely the goal of 
$p$-Adic Physics [11] to remove the notion of spacetime per se and replace it by objects and their relationships. A truly {\bf Categorical } view of the Universe. 
An extension of Einstein's motion Relativity and Nottale's Scale Relativity into a unfied Scale-Motion Realtivity was outlined
briefly in [13].      
Whatever the "final"   view of the world may be, it seems that it is wrong to assume that Quantum Spacetime has a fixed 
dimension. On the contrary, it may have uncountably-infinite dimensions
as El Naschie has argued [4]. Taking this infinite-dimensional point of view 
allows us to eliminate the notion of a {\bf EPR }, and possibly, Black-Hole Information Loss "paradoxes ". For this reason we believe it ought to
be investigated further. Dimensions are not {\bf fixed} absolutes. They are {\bf resolution} dependent concepts.  

Quantum Gravity is {\bf not} a quantization of the spacetime coordinates, metric.....If this were the case, one would have had quantized the spacetime 
coordinates long ago. In String Theory, from the two-dim world sheet point of view , the spacetime coordinates are nothing but a 
finite number of scalar fields whose quantization is 
essentially trivial by selecting the conformal or orthonormal gauge. The same arguments applies with the ( linearized )  spin two graviton. 
Quantum Gravity 
it is something much deeper than the naive notion of coordinates and gravitons. It is something that doesn't need any 
spacetime background nor metrics whatsoever. Morever, it involves something that disposes of the ill-conceived notion of having a {\bf fixed} dimension.  
The classical spacetime that we perceive with our senses is just a long distance averaging 
effect associated with a 
quantum network of processeses
of a deeper underlying Quantum Universe. Einstein's Gravity 
is an effective theory as suspected long ago. To merge Quantum Mechanics with Relativity it is necessary to enlarge the Einsteinian view of Relativity to a 
New Relativity Principle [1]. To proceed further one has to demolish the concept of dimension as an {\bf absolute} , as an idol.    

To sum up what has been said so far : 

The New Relativity Theory forces upon us to take a radically different view of the Quantum World,  an ultimate Machian/Zenthropic view,  and to 
dismiss the concepts of false absolutes (idols) of dimensions, spacetime, cosmological constant , from our classical minds, as Finkelstein [8]  has advocated.
If a true {\bf evolution} ( revolution) 
of Physics is to take place 
one has to embrace the plausible
extensions of Relativity as Finkelstein has insisted [8]. 
For those who believe that we have reached  the end of the road, the end of Physics, 
we feel that they are setting 
themselves for similar surprises  that Lord Kelvin experienced  with the advent of Quantum Mechanics and Relativity. 

To this day , to the best of our knowledge, there is no satisfactory definition yet of Quantum Field Theory. 
QFT today is being challenged by deeper concepts :  Noncommutative Geometry, Quantum Groups, Hopf algebras, Monoidal Braided Categories, Braided QFT, etc.... 
Relativity itself is hereby extended to a deeper 
meaning by the New Relativity Theory : Scale Relativity and Cantorian-Fractal Geometry. 
A Nested Hierarchy  of Histories have replaced the old fashioned concepts of spacetime events; 
vectors and tensors have been replaced by Clifford-multivectors; 
Riemannian Geometry by Finsler Geometry  and by Fractal-Cantorian, 
Non-Archimedean, $p-Adic$ , Noncommutative and Nonassociative Geometries..... 

Recently we have proposed to even abandon the the idea of the cosmological constant as a constant. The so-called cosmological constant is {\bf not} a constant in its 
definition ! It is observer-dependent within the framework of the New Relativity Theory [1]. Trying to estimate the absolute values of such a " constant " 
is like trying to detect absolute spacetime motion and to  verify the existence of 
the ether !  Such ideas that the vacuum energy could be 
observer dependent orginated with discussions held in Trieste by one of us with Miguel Cardenas and Devashis Benarjee [9]. Even the notion of the "vacuum" per se !
Special  Relativity demolished such framework of thinking. We believe that the New Relativity Theory will also replace the existence of such 
ill-conceived notions that spacetime has a fixed dimension and that the cosmological constant has an well defined absolute value in all frame of references.  

The observed spacetime dimension of $D=4$ is interpreted in this New Relativity Theory [4] as a result of an  {\bf averaging procedure } 
over all the possible infinite values of
Quantum Spacetime. In a sense it  is similar to what happens with the statistical distribution of velocities of a gas. There is an average velocity 
( average over all the infinite possible values of the statistical ensemble ) proportional to the Temperature.  
To assume that there is a spacetime compactification from {\bf D=11}  to {\bf D=4} ( like it is assumed in mainstream Physics today )   
is an incongrous assumption in this New Relativity Theory : 
it is like saying that there is a " velocity compactification/decompactification  " from higher/lower  velocities to the 
average observed velocity in a gas. Problems with the compactification picture of Superstring theory from  {\bf 10}  to {\bf 4}  were already alarming signals 
when billions and billions of possible
four-dimensional theories of the world were obtained : the so-called uniqueness of string theory went out the window when this was found. String theory wasn't the problem, 
assuming a fixed dimensions was ! 
The fact that there might be an  Statistical approach to the Dimensions, and to Quantum Gravity per se,  was already lurking behind the scenes long ago 
in the work of Hawking : Black Hole Thermodynamics !

\bigskip

\centerline{\bf 3.  There is No  Such Thing as an Einstein-Rosen-Podolski Paradox in the New Relativity} 

\bigskip 

We will present a few-lines proof why there is no {\bf EPR}  Paradox within the framework of the New Relativity Theory 
if one assumes that information flows in a similar fashion as 
ordinary charges in Electromagnetism ; i.e information is to be thought of as a " field " [14]. 
Interestingly
enough, this will be our {\bf only} assumption. We are 
{\bf not} implying that there is such a thing as a " fifth " force in Nature found one morning in the closet of our homes after a bad night. 
We are just voicing out what has been 
{\bf irrefutably} proven over and over by experiments.   

Take an electron-positron pair colliding at the center {\bf O} of an infinite dimensional sphere, $S_D$ for $D\rightarrow \infty$, 
at a givent moment we call $t=0$. After the collision a pair of two photons will travel in opposite directions imposed by energy-momentum conservation. 
 An any given moment after the collision, we can locate those two 
photons at the surface of a multidimensional sphere of radius $R =ct$. The flux of information from the center of the sphere {\bf O} 
flowing from the moment of the $e^-/e^+$ collision radially outwards through the hypersurface is : 

$$\Phi = \oint {\vec J}_D .d{\vec S} _{D-1} = {J}_D S_{D-1}. \eqno (2)$$
This is nothing but the usual Gauss Law in Electromagnetism. The $D$-dimensional information-current, $J_D$,  points radially outwards from the center {\bf O}. 
Due to hyper-spherical symmetry its magnitude only depends on the radius $R =ct$. At each  given point on the hypersurface, the current is pointing radially outwards 
 and has the same value of magnitude, $J_D (R)$,  along all the points of the  hyper-sphere, This  
is why one can pull out the current outside the integral. The hypersurface $S_{D-1} $  encloses inside a $V_D$ volume given in terms of gamma functions. 
Similar considerations apply to the higher-dimensional solid angle : 

$$V_D = { \pi^{D/2} R^D \over \Gamma ({D+2 \over 2} )  }. ~~~S_{D-1} = {dV_D \over dR} = R^{D-1}\Omega_{D-1}.~~~\Omega_{D-1} = {1\over R^{D-1} } {dV_D \over dR}. \eqno (3) $$

Therefore, the total information-flux is given by the usual Gauss Law :

$$\Phi = J_D(R).R^{D-1}  \Omega_{D-1} =  J_D (R).R^{D-1}. {1\over R^{D-1} } {dV_D \over dR}=  J_D(R).{ D \pi^{D/2} R^{D-1} \over \Gamma ({D+2 \over 2} ) }. \eqno (4) $$  
Now we take the $D \rightarrow \infty$ limit and make use of Stirling's asymptotic formula for the gamma function :
( Pictures drawn on a  Mathematica package also  verify explicitly the results below ) 

$$lim_{D\rightarrow \infty}~\Gamma ( {D+2 \over 2} )  \sim \sqrt {2\pi} ({D+2 \over 2}) ^{ {D+2 \over 2}} e^{- {D+2 \over 2}} . \eqno (5)$$
By Radius $R =ct$ one means radius in Planck scale units. We will  set the Planck scale to {\bf 1}. So by $lnR $ in all of the formulae below we mean $ln (R/\Lambda) $  
Otherwise the units will not match up.  
 
As $D\rightarrow \infty$ one can verify that in the asymptotic $D=\infty$ limit the numerator expression for the flux approaches :

$$exp~[ ln D +{D\over 2} ln \pi + (D-1)  ln R ]\sim exp~[ ln D + {D\over 2}  ln \pi + D  ln R ] \sim exp~ [ ln D + D ln \pi + D ln R ]  \eqno ( 6 ) $$
whereas the denominator approaches :  

$$exp~[ {D+2 \over 2} ln ({D+2 \over 2}) - {D+2 \over 2} ] \sim   exp~ [ D ln D -D ] \sim exp~[ (D-1) ln D ] \sim exp ~ [ D ln D ]  .\eqno (7)$$
Hence,  the flux in the infinite $D$ limit is :

$$\Phi = J exp~ [ ln D + D ln \pi + D ln R -D ln D ] = J e^ {\alpha}. \eqno (8a)$$
To be precise, upon reinserting the Planck scale one has that the flux is given in Planck units as : 
$$\Phi = J (\Lambda)^ {D-1} e^\alpha = Je^\alpha \times 1^{D-1} = Je^\alpha  . \eqno (8b)$$ 
where $\alpha $ is :

$$\alpha = D ( ln \pi + ln R ) + (1 - D ) ln D \sim  D [ ln \pi +ln R - ln D ]  \eqno (9) $$

For $finite $ times , in units of $\Lambda =1$, $ R =ct \not= \infty $ the coefficient $\alpha $ goes to {\bf negative} infinity :     

$$\alpha \sim -D ln D \rightarrow - \infty  \Rightarrow e^{\alpha} \rightarrow 0  \eqno (10)$$
So

$$ lim_{D \rightarrow \infty} ~ J_D (R).{ D \pi^{D/2} R^{D-1} \over \Gamma ( {D+2 \over 2} )  }\rightarrow  J_\infty (R)\times 0 = \Phi  
\Rightarrow J_\infty (R)\rightarrow \infty . \eqno (11) $$
hence, in the $D=\infty$ limit, the current ( in Planck units, $\Lambda =1$ ) blows up. This is {\bf not} because there is a {\bf superluminal} speed of information. It is because the 
hyper-volume, hyper-area elements, for finite values of $R$, go to {\bf zero} in infinite dimensions!. 
Everything shrinks to a {\bf hyperpoint} despite the fact that the radius is {\bf not} zero ! The {\bf hyperpoint} is the infinite-dimensional version of a point in ordinary finite-dimensional spacetime. 

The current is as usual of the form : $J=\rho v$. 
As the hyper-volume, hyper-area elements, for finite values of $R$,
go to zero, the information charge density $\rho$, charge per unit hyper-volume, blows up !. 
The information charge density diverges at the hyperpoint. The information velocity $v$ is constant and cannot exceed the speed of light.
 From the point of view of an infinite-dimensional observer, {\bf all } the points of the hypersurface are {\bf interconnected}. There is no such thing as 
{ \bf non-locality} in Quantum Mechanics. This is an illusion due to the shrinking to {\bf zero} ( for finite radius)  of the infinite-dimensional volume of the hypersphere, 
resulting from the asymptotic behaviour of the gamma functions !

This corroborates Mach's  brilliant insight that everything is connected in the ( Quantum ) Universe. 
What happens here and now, affects everything in the Universe in an instant. 
Based on the recent teletransportation experiments of a single photon by several experimental teams, this
view of the Quantum Universe may lead an advance future generation of open minded scientists to achieve  the ultimate communication system : 
instant exchange of information to anyplace in the Universe by tapping into the infinite dimensions of Quantum Spacetime. 
An speculative application  of this would be to 
tele-transport a quantum copy of 
the human genome to other distant Planets in the Universe suitable for life. This would be a way out of the Galactic bounds we live in and 
an escape of the ultimate fate of the earth : consumed  by the Sun when it becomes a Red Giant.    

Spacetime travel in an instant  will be much harder to achieve if by travel on means tele-transporting a quantum copy of ourselves to another point in the Universe.
In order to do that one has to be able to tele-transport our consciousness as well. We adscribe to Penrose's view that consciousness is a non-algorithmic process. 
This agrees with the Uncountably-infinite number of dimensions of the Cantorian-Fractal Spacetime view of El Naschie [4]. 
It would be impossible for a Quantum Turing Machine ( a Quantum Computer) to 
quantum-process such vast of uncountably-infinite number of quantum bits. Never, in our wildest dreams we could possibly count such large number of dimensions of the 
Cantorian-Fractal Spacetime of El Naschie [4] . Such World is not a mere Mathematical abstraction : it is {\bf essential}  
for Consciousness to emerge. It is desirable that 
The Theory of " Everyhthing " 
should include Consciousnes. The Theory of Everything has to account for the existence of 
Conscious life and when, why, how, and {\bf for what} it 
emerged from the Quantum Universe. A "pointeless " Universe is another one of those alarming signals that something 
is inherently incomplete with our view of the World. We believe that it
is not sufficient
to dismiss these questions as "meaningless metaphyiscs ".

Upon closer inspection of 
eq-(11), if one were to set  $J=finite$ ; this would imply that the information flux $\Phi=0$ so by Gauss Law there is $no~net$ 
information charge enclosed in the hypersphere. This is not correct for the following reason. 

Nottale's Scale Relativity implies that it is not possible to have zero measures with {\bf zero} resolutions. 
It is possible to have zero measures but with ( nonzero) Planck scale resolutions. The $e^-/e^+$ pair never goes  beyond the minimum Plank scale resolution. 
The center {\bf O} of the hypersphere is not a physical point. It is a smeared fuzzy hypersphere of infinite dimensions but with a nonzero Planck scale radius. 
This is a reason why Noncommutative Geometry, Fuzzy Phyiscs, Quantum Groups ....could be the right approaches to look at the world at small scales. 
Thus the information charge 
is distributed  " uniformly " , in discrete bits of Planck hyper-area, in Planck units ,   over the outer " surface " of the 
hyperball  of Planck radius. There is {\bf no} inside. Inside is meaningless notion below the Planck scale, 
this is why the information charge has to reside on the " surface ". It would not be so surprising if this mechanism could be linked to the Bekenstein-Hawking 
entropy-area relationship of Black Holes.
  
The number of dimensions increases as one probes finer and finer $resolution$-scales ( not to be confused with lengths, although they both have the same units). 
By resolutions one means the resolutions that a physical apparatus can resolve. 
Resolutions which are not the same thing as the spacetime labels of a " point ",
event " like $x^\mu$. 
Resolutions that so far 
( until Nottale) have been overlooked in the description of Physics.     
As one approaches asymptotically the Planck scale $resolution$ , 
the hypersphere of Planck scale radius becomes more and more " visible" to us . To be able to reach this 
limiting " threshold " of $resolutions$ in our physical apparatus, an infinite 
amount Energy is required as Nottale has argued. By the same token that it takes an infinite amount of energy to accelerate a mass ( nonzero rest mass) from rest to the 
speed of light, it takes an infinite energy to probe 
Planck scale-resolutions. The 
final  
infinite-dimensional hypersphere, containing the information charge located at the " origin " {\bf O}, shrinks to a hyperpoint of zero size , but finite 
Planck radius, The information charge density also  {\bf diverges} at the infinite-dimensional point : the hyperpoint of nonzero Planck radius. 
Exactly in the same way it did for hyperspheres of radius $R=ct$ upon taking the infinite dimension limit.

Concluding, the flux $\Phi$ is {\bf not} zero. There is a net information charge enclosed by the hypersphere with center {\bf O} and radius $R=ct$.

Exactly the same argument occur if one asks the question : What does one of the  photons  " see " ? 
It will " see "  the other photon at a distance 
$l=2R=2ct$ ( in Planck units) move away at the speed of light. Due to the Doppler effect, the frecuency will be redshifted so much that the photon will appear to 
be completely dark, with zero frecuency .  
For finite values of the radius, $l=2R=2ct$, the hypersphere centered at one of the photons will again shrink to a zero size , 
to a hyperpoint, in the infinite Dimensional ( large $D$ ) limit.
Therefore, when a Macroscopic Observer with a Physical Apparatus   measures the polarization of one particular photon, 
it will transfer its information to the other photon in an instant  
due to the fact that both of the photons have access to an extremely large number of dimensions in comparison to the macroscopic observers; i.e 
the photons truly live inside the hyperpoint. For this reason, they are able to exchange information in an 
instant without actually having 
a superluminal speed of information ! It is the information charge density ( and information current $J$)  that diverges once again 
at the hyperpoint, and not the information speed. 
It is only an illusion due to the shrinking to zero of the hypersphere in the infinite dimensional limit.

Of course, the $e^-/e^+$ pair does  not attain such infinite energies to probe Planck scale resolutions, 
they come very close to each other but never reach the Planck scale. 
As they approach each other more and more dimensions become visible to them. Much more dimensions than the 
dimensions of the apparent one-dimensional world to a macroscopic observer looking at the line   between the two emerging photons while performing his experiment. . 
Effectively, the number of dimensions of the world visible to the $e^-/e^+$ pair, and the two emerging photons , is 
very high in comparison to the apparent $D=1, D=4$ of the macrospcopic observers , that for all practical purpopses, one can take the 
infinite dimensional  limit of the gamma functions. 
The diagrams explicitly
show that the hypervolume, hyperareas fall-off  very rapidly to zero as $D$ moves far away from the $D=4$. 
It is not necessary to actually take the infinite dimensional  limit too literate. 
    
A related textbook issue is the following : Imagine a rapid moving observer passing by ourselves during the night while we are gazing at the stars. Due to the Lorentz 
contraction the celestial sphere that he experiences will naturally shrink with respect to us. 
It shrinks, but does it appear {\bf flattened } ? The answer is no. 
One can view the Lorentz transformations in spinor terms as a $SL(2, C ) $  Mobius transformation. Since the Mobius transformation maps circles to circles, 
the celestial sphere
will have shrunk in radius only but it will {\bf not}  be flattened. 

Similar analogy happens to the photon. What does a photon " see " ? Since we have said earlier that one cannot for certain answer such questions. We can only follow 
what we know so far : 
Due to the infinite Lorentz 
contraction the celestial sphere will shrink to a point. Doesn't this contradict Nottale view that the Planck scale is the mimimal length ? 
The answer is no. Once again we have to take the variable dimensions of the Quantum World that a photon experiences. 
The photon is a quantum entity. Nobody can deny this. 
The photon of a given energy $E=\hbar \omega$ will 
probe resolutions larger than the Planck scale. Rigorously speaking , we should write : $E =\hbar_{eff} (k^2) \omega$. 
In [1] we have shown that 
the New Relativity Theory demands an energy-dependent effective Planck constant so that 
$[x,p] =i\hbar_{eff}(k^2)$ to reproduce the full blown Quantum Spacetime Uncertainty Relations that are more general than the 
String Uncertainty Relations : we have included the effects of {\bf all} extended objects [1]. 
 It was shown
rigorously why one cannot 
probe resolutions smaller than the
Planck scale. As energy begins to be pump-in, one cannot probe smaller scales. Spacetime actually starts to grow. It is 
possible that a polymerization growth 
process of the Quantum Spacetime begins : an infinite chain of self similar branched polymers is triggered and baby universes branch off. 
The Quantum Universe might be an ever self-reproducing , self-recursive, self-iterated fractal process as Linde has suggested.

Only at infinite energy will a photon be able to probe the Planck scale. 
The celestial hypersphere that the photon 
" sees "  has a radius of the order of the inverse photon Energy, roughly, assuming it is a low energy photon, 
Energy and resolution are inversely correlated at that level., not at higher enegy levels. Scale Relativity implies that the Compton wavelength and momentum 
are {\bf decoupled} as one approaches Planck scales. It takes an infinite energy to probe the Planck scale. The Planck scale is the ultimate Ultaviolet Regulator. 
However, due to 
the effectively large number of dimensions that the photon has access to, 
despite the fact the hypersphere  has a nonzero radius, the celestial hypersphere shrinks to {\bf zero} size consistent with the infinite Lorentz contraction ! .
It is true that one has to construct the full Scale-Motion Relativity [13] to be fully rigourous and consistent. 
We have presented a solution to the apparent paradox of how one can have a zero measure/size  ( due to the infinite Lorentz contraction) with 
a nonzero resolution for a radius : Infinite ( large number of  )  Dimensions is the key once again !

Therefore, in essence : By introducing the notion of {\bf hyperpoint} in physics, which is forced upon us by the New Relativity Principle 
as a result of having a truly infinite dimensional Quantum World. we have imbued a mathematical point with a true {\bf physical}  meaning : 
it is an infinite-dimensional hypersphere, of {\bf zero} size but {\bf nonzero}  radius !.

When $t =\infty$ then the coefficient $\alpha$ will {\bf no}  longer be negative infinity due to a cancellation  between $ln R$ and $  ln D$ : 

$$ \alpha \sim D [ ln \pi +ln R - ln D ] =  D [ ln \pi +ln (ct)  - ln D ] \sim D ln \pi \Rightarrow e^\alpha \rightarrow \infty . \eqno (12) $$ 
In this case, one has the opposite result :  the value of the information current $J_D $ at $R =\infty$ collapses to zero, as it should on physical grounds. 
The information field must vanish at infinity in {\bf any} dimension, finite or infinite. As the photons move away from eachother, if one waits  an infinite
amount of time to peform the EPR experiment, the photons will {\bf no}  longer be {\bf correlated } !

To sum up : 

The EPR Paradox only occurs to the one-dimensional beings ( or finite-dimensional beings) living along the linear path ( around the linear  path) 
of the photons who wish to perform the EPR gedanken experiment.
From their finite-dimensional point of a view, QM appears to be {\bf non-local } : a superluminal transfer of information appears to take place. 
From the point of view of the New Relativity Theory there is no paradox because Quantum Spacetime is truly infinite-dimensional. 
For those Quantum-dimensional beings who  
were able to tap into  the effectively "infinite" number of  dimensions of Quantum Spacetime at the very " moments "  when the $e^-/e^+$ pair collided, 
at a very small distance separation among them, 
distance which cannot be smaller than the Planck scale as indicated by Nottale's Scale Relativity,  there is no 
such Paradox at all ! : the information current blows up because from their  infinite-dimensional point of view, for {\bf finite } values of the radius, 
the hypersphere has shrunk to a hyperpoint. The transfer of information to the two photons, about the spin and other quantum numbers of the $e^-/e^+$ pair, 
occurs in an instant ! 
Every point in their universe is inter-connected as Mach argued long ago.  
Similar arguments apply to the two photons when a macroscopic observer measures the polarization of one photon, the information is transfered  
to the other photon in an instant 
via an effectively " infinite "  dimensional ( relative to the macroscopic observers) Quantum Spacetime accesible to them.   

This should encourage us to view Feynman's path integral formulation of QM taking {\bf all} posible paths in a {\bf finite} dimensional spacetime, 
from the New Relativity Theory point of view : 
it is possible to have a finite number of paths in  
an Infinite Dimensional Quantum Spacetime.  The main question is : 
Where does the Feynman statistical complex-weighting of the paths via the $e^{iS}$  comes from ? 

The partial answer was given by Ord [10] , Nottale [3] and others : 

Since fractal paths have a dominant weight in the path integral compared  to the smooth ones, the latter have a zero measure compare to the former, 
roughly speaking, Quantum Effects manifest or channel themselves via the fractality of spacetime. Although there are people who do not subscribe to this view.  
Fractal curves 
are continous but nowhere differentiable. 
This means that the derivatives are {\bf discrete}-valued. The discrete 
jumps of the values of the tangents are "quantized" in units of 
what has been called by mathematicians the " Planck " constant of a curve. In this fashion the Feynamn $e^{iS}$ weighting factor is interpreted 
although , we must say 
that no rigorous proof of this
has been given as far 
as we know.

Fractals and Scale Relativity are essential because as the resolutions that a physical apparatus can resolve reach the mimimal Planck scale resolution 
( resolutions must not to be confused with statistical uncertainties nor with ordinary lengths)  
the number of fractal dimensions blows up. For a {\bf new}  Phase space path integral derivation of Feynman's particle 
propagator that is $roughly$   based on these ideas that a fractal particle " path " 
can have a meaning in QM see  [7].

The apparent superluminal information velocity happens in other aspects of Physics.    
There  is a very simple analogy with superluminal jets in Astrophysics [5] . If one takes a flash light at a sufficiently large distance from a wall and rotates
it very rapidly , the image on the wall can  appear to move faster than light. However the image is {\bf not} a truly physical object. The physical photons never
move faster than light. The image is comprised of many different photons and not of a fixed particular  number of them . 
The maximum angular velocity of rotation of the flash light is bounded by Special Relativity :

$$ \omega_{max} = {c \over r} \eqno (13)$$ 
where $r$ is the length of the flashlight. If the distance to the wall is $R$ then the apparent velocity of the shadow is :

$$v = {cR\over r} > c.  \eqno (14)$$
The ( unphysical object) shadow can move faster than light. One does not even have to go to such extremes of achieving the maximal angular speed  
for the flash light , one can simply
choose the wall far enough, and the flashlight sufficently bright, ( $R$ large enough )  so the image on the wall moves with a superluminal  velocity $ \omega R > c$. 
Exactly similar arguments occur with the phase velocity in wave propagation. The phase velocity
can be greater that $c$ but the physical group velocity is always bounded by $c$. Taking the number of Dimensions to infinity, mimics this simple example of taking the 
distance to the wall far enough and rotating the flash light fast enough.

Similar arguments can  be taken with the so called Back Hole Information loss Paradox. Since Quantum Spacetime is truly infinite dimensional, there is no 
such thing as an Information Loss. This information is stored in {\bf all} the infinite number of dimensions that are inaccesible to an  outside low energy observer. 
There is information radiated away and a remnant " hidden " in the infinite number of dimensions inaccesible to the outside observers. Black Hole evaporation 
{\bf stops}  at the minimal scale in Nature : the Planck scale, reaching a maximum temperature, Planck's Temperature. Scale Realtivity not only induces an effective 
value of the Planck's constant : $\hbar_{eff} (k^2)$ [1], it also affects the Boltzmann constant as well : $k_B (k^2)$ so that :
$k_B (k^2) T = \hbar_{eff} (k^2)\omega$. As one reaches 
the Planck scale, energy blows up but the temperature reaches asymptotically the maximum Planck Temperature ( thermal Relativity). One must have a standard of temperature to compare temperatures with. That maximum universal standard is the Planck Temperature whose definition in $D=4$ is : 

$$T_P = \sqrt { {\hbar c^5 \over G k_B^2 } } = 1.42\times 10^{32} K . \eqno (15)$$

Astrophysicits have been baffled by recent 
findings that there are unexplained 
extremely bright and 
unrelenting sources of energy. It is warranted to study these phenomena within the framework of the New Relativity Theory. 
To be able to " see " all the information one has to tap into all the infinite number of 
dimensions of the Quantum Spacetime. To achieve that one requires infinite amount of energy to probe the Planck scale resolutions 
according to the Scale Relativity Principle . 
At that scale (infinite)  Dimensions, (infinite) Energy and (infinite) Information merges into the " Omega " hyperpoint , the " Trinity " hyperpoint.....
the ultimate infinite-dimensional point :    
At that scale, the " Trinity " hyperpoint, Dimensions, Energy and Information are indistinguishable from each other. More details will be given later.

\smallskip

\centerline{\bf Acknowledgements}
\smallskip 

We are indebted to E. Spallucci for a very constructive critical remarks.  
We thank G. Chapline. L.Nottale , W. Pezzaglia, M. El Naschie and D. Finkelstein for 
illuminating discussions. Finally many thanks to C. Handy and M. Handy for their  assistance and encouragement . 
\smallskip  

\centerline{\bf References}

1. C. Castro : " The String Uncertainty Relations follow from the New Relativity Principle " 

hep-th/0001023. " 

" Hints of a New Relativity Principle from $p$-brane Quantum Mechanics " hep-th/9912113.

" Is Quantum Spacetime Infinite Dimensional ? hep-th/0001134. 

"Towards the Search for the Origins of $M$ Theory, Loop Quantum Mechanics and 

Bulk/Boundary Duality ........hep-th/9809102.

2. W. Pezzaglia : " Dimensionally Democratic Calculus and Principles of 
Polydimensional

Physics " gr-qc/9912025.

3. L. Nottale : Fractal Spacetime and Microphysics, Towards the Theory 
of Scale Relativity

World Scientific 1992.

L. Nottale : La Relativite dans Tous ses Etats. Hachette Literature. 
Paris. 1999.

4. M. El Naschie : Jour. Chaos, Solitons and Fractals {\bf vol 10} nos. 
2-3 (1999) 567.

5. Phillip Morrison : Conversations held with Carlos Castro at MIT in 1980. 

6. . C. Castro, A. Granik et al : In preparation. 

7. S. Ansoldi, A. Aurilia and E. Spallucci : Eur. J. Physcs C {\bf 21} (2000) 1-12. quant-ph/9910074.

S.Ansoldi, C. Castro, E. Spallucci : Class. Quantum. Gravity { \bf 16} (1999) 1833.  

8. D. Finkelstein : " Third Relativity " Georgia Tech preprint, January 2000.

" Emptiness and Relativity " Georgia Tech preprint. December 1999. 

9. D. Benarjee, M. Cardenas : Private Communication.

10. G. Ord : J. Chaos, Solitons and Fractals {\bf 10} (2-3) (1999) 499. 

11. M. Altaisky, B. Sidharth : Journal of Chaos, Solitons and Fractals {\bf vol 10} (2-3) (1999) 167.

l. Brekke, P. Freund : Phys. Reports {\bf 231} (1993) 1-66.

V. Valdimorov, I. Volovich, E. Zelenov : $p$-adics in Mathematical Physics. World Scientific 1992. 

A. Khrennikov : Non Archimedean Analyis, Quantum Paradoxes , Dynamical Systems and 

Biological Models. Kluwer Publisng 1998. 

12. L. Nottale : Private Communication

13. C. Castro : J. Chaos, Solitons and Fractals {\bf 10} (2-3) (1999) 295. 

14. M. El Naschie : On the Unification of the Fundamental Forces and Complex Time

in the ${\cal E}^{(\infty)}$ Space. Jour. Chaos. Solitons and Fractals {\bf 11} (2000) 1149-1162. 

\bye